# Just in Time to Flip Your Classroom
Nathaniel Lasry, Michael Dugdale & Elizabeth Charles

With advocates like Sal Khan and Bill Gates[1], flipped classrooms are attracting an increasing amount of media and research attention[2]. We had heard Khan's TED talk and were aware of the concept of inverted pedagogies in general. Yet, it really hit home when we accidentally flipped our classroom. Our objective was to better prepare our students for class. We set out to effectively move some of our course content outside of class and decided to tweak the Just-in-Time-Teaching approach (JiTT)[3]. To our surprise, this tweak - which we like to call the flip-JiTT - ended up completely flipping our classroom. What follows is narrative of our experience and a procedure that any teacher can use to extend JiTT to a flipped classroom.

## What are flipped classrooms?

Flipped classrooms invert the conventional way we teach. A simple description (from www.flippedlearning.org) is: "*Flipped Learning occurs when direct instruction is moved from the group teaching space to the individual learning environment.*" In traditional classrooms, a teacher who knows the content presents it to students who do not know it. Thus, in class the focus is on presenting and transferring knowledge to students. In science courses, this usually means that the students' first exposure to the material is in the lecture hall. Outside of class, students are given 'homework', such as problem sets or exercises, that help them make meaning from lecture materials. In contrast, students in flipped classrooms are required to gather information on their own *before* they come to class. One possibility for moving the instruction to the "*individual learning environment*" is taping lectures, placing them online and assigning them to students before they come to class. However, there is more than one medium that students can use to gather information before coming to class. Students can be assigned readings or referred to online resources such as websites, videos and simulations. The objective is to move the information transfer outside of the classroom, not to have students understand all the content before coming to class. So what happens in class?

Teachers' roles during class time change in flipped classrooms. Instead of focusing on presenting information, teachers focus on the significant gaps that students may have in their understanding. Teachers use subject-matter and pedagogical expertise in class to help students make meaning of the information they gathered before class. Teachers help students create connections between new and prior knowledge, usually by giving more complex assignments in class; much like the kind of exercise that traditionally would have been given as homework. Hence the term 'flipped' : what is usually seen as homework is now classwork, while the traditional classwork, is now done as homework.

**Why Flip?**

Did you ever think about what the most expensive resource in a classroom was? More than the computer, projector or digital-blackboard combined? Yes, the teacher. Now, does it make sense to use the most expensive resource as a book… when you already have a book?

The idea of the flipped classroom is simple. Teachers have expertise. Expertise is more than the quantity of facts and concepts they know. It is better described by the way they connect these into coherent and meaningful conceptual structures that they know when and how to use. The role of teachers in flipped classrooms is better aligned with their expertise. Instead of presenting information, teachers help students connect the information they gathered before class into meaningful chunks. Teachers help students overcome their conceptual difficulties and help students recognize when and how to apply the newly constructed knowledge.

**Our Accidental Experience**

The change required to flip your classroom is not trivial. As experienced physics teachers, we were intrigued, but not prepared to make big changes. What we were prepared for was another semester of a college course on Electricity & Magnetism (E&M) using Peer Instruction[4-6]. We had a few issues we wanted to iron out. In our experience, students spent a fair amount of class time discussing abstract E&M concepts with each other. Using Peer Instruction meant that we had organized our classes around short lectures that were followed by conceptual questions. Students answered these questions individually before discussing them with their peers.[4-9] This time spent pairing and sharing in class meant that there were some topics we could no longer cover in class. Students would have to cover these topics on their own, outside of class. As teachers in a multi-section course (we each taught 1 of 10 sections of the course), our students would have to write the same exam as all other students registered in the course. We had to find a way to get students to be responsible for the material we would no longer cover because of the class time spent in teacher-facilitated peer-discussions. In cases like these, Mazur[6, 7] had proposed using Just-in-Time-Teaching (JiTT)[3].

The JiTT approach is an ideal companion to Peer Instruction because it is a structured approach that helps student prepare for class. In JiTT, readings or other information gathering activities are assigned *before* a topic is seen in class. Students then complete an online assignment that checks whether they engaged in the preparation activity and asks what they find difficult or confusing. For instance, students can be given a reading quiz or a couple of conceptual questions to find out if they read carefully. The central feature of JiTT is the feedback question that follows. The standard form of question is a variation on the theme: "what did you find difficult in the readings?" The instructor receives student feedback in the form of responses to this question a number of hours before class (often the night before) and reviews it 'Just in Time' for class. Each class begins with (and can be designed around) what students find difficult. By being exposed to the material before coming to class, students are more deeply engaged in the process of their own learning, and are better prepared for an active learning environment. This alleviates some of the time pressures that teachers face in covering content and allows the instructor to focus on making deeper connections between concepts.

We had tried JiTT in the past with varying levels of success. The main issue was getting students engaged with the material before coming to class. Our, albeit anecdotal experience, was that those who read were seldom sufficiently engaged in their reading. Our objective for the current semester was simple: We wanted students to come to class prepared. We set out to create a structure that would make JiTT easier for us to use and harder for students not to use. Following the JiTT approach, we wanted to monitor students' progression before they came to class, find out what they understood and what they had difficulty with. So we tweaked the standard JiTT approach. What happened took us by surprise. By tweaking JiTT, we accidentally flipped our classroom…

**What We Did: The Flip-JiTT**

We used LON-CAPA (http://www.lon-capa.org) as a course management system to find out what our students knew and monitor what they were doing *before* they came to class. We chose LON-CAPA because it is an open-source platform with lots of physics content. However, this structure can be used on any course management system. We tweaked the standard JiTT procedure as follows:

1- **What do you know?** We began by asking students to reflect and state what they knew about a given topic.
    a. The first statement always was: "Before you start (…) it's important to establish what you already know about the topic"
    b. This was followed by topic-specific statements for each class, such as: "We've all experienced electrostatic forces, whether we realized it or not. In the space below take 2 or 3 minutes to write 3-5 short sentences on what you already know about electrostatic forces, where you've observed them and how they behave."

2- **Gathering Information :**
    a. Readings from the textbook were assigned.
    b. We also assigned links to relevant videos, websites and simulations.
        i. For instance, we made frequent use of PhET simulations and regularly pointed students to lectures by Walter Lewin at MIT (most often telling students which time intervals in a given lecture were most relevant). The selection of these online resources can be very time intensive, and represents the bulk of the preparation time of this approach.

3- **"Warm-up" problems**:
    a. We typically assigned 4 to 6 LON-CAPA questions. These questions ranged from simple single-concept computational questions, to more difficult non-numerical conceptual questions to mildly complex questions that meshed conceptual and computation aspects.
    b. Students were not expected to have understood all the material before class. We always gave them at least 5 tries, with no penalty for getting it wrong. However, they could get an unlimited amount of tries if they came to see their teacher. In the past, we had observed students trying a problem 30 or 40 times before they gave up. We put a cap at 5 tries and told them they could get more if they came to see us with questions. We also gave at least 48hrs to complete the "warm-ups", provided those 48hrs would overlap with our office hours.

4- *Now* **what do you understand?**
    a. For instance, the statement at the end of the electrostatic force assignment was: "Now that you had a chance to read the text and work a few problems your understanding of electrostatic forces might have changed a little (or a lot!). Take another 2-3 minutes to write 3-5 short sentences on what you now understand."

5- **Reflect on what you learned**.
    a. We displayed the latest entry of what they stated they *now* understand side-by-side with their initial statement. In our electrostatic warm-up, this was framed as follows:
        i. "Before you started you wrote the following about the electrostatic forces:" [initial statement]
        ii. "At the end you wrote;" [latest statement]
        iii. "Take a few minutes to reflect on what you learned in this exercise. Write a short paragraph (5-6 sentences) on how your understanding has evolved."

6- **What do you still find hard or confusing?**
    a. Students were asked: "What questions, if any, do you still have about the material covered in this "warm-up"?
    b. What areas would you like to cover more deeply in class?
    c. Is there anything you still find confusing?
    d. If not please state what you found to be the most interesting.
        i. This last item was added to make sure that students always write something. If not, there is too great of an incentive to write : "I found nothing confusing".

**Figure1** The flow of our flip-JiTT side-by-side with conventional JiTT.

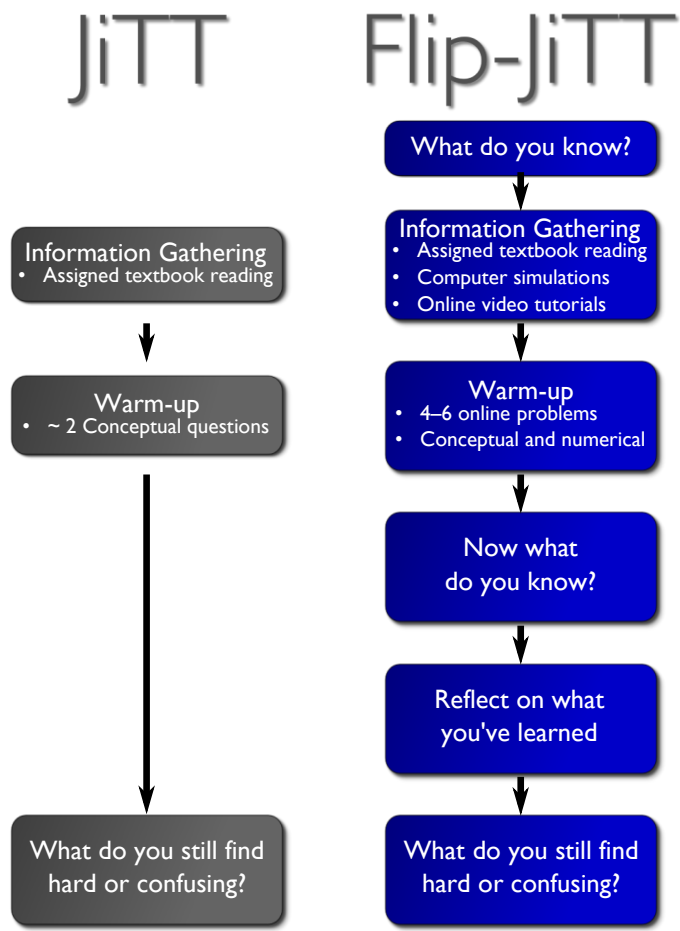

**What We Found Using the Flip-JiTT**

Using these LON-CAPA warm-ups, we could now track what our students were doing before class. Grades were assigned to these warm-ups, so students had incentives to participate. Tracking student participation, we found a compliance rate of 83% (we assigned 11 warm-ups to 69 students; grades were assigned to 631 out of a total possible of 759 assignments for a participation rate of 83.1%). We no longer needed to find out if students had read because students had to solve the warm-up questions. Whether they read or not, they had to gather enough information to understand the problems and solve them. They might not have read the textbook. Instead, they may have listened to an online lecture, looked at a website or played with simulations. One thing we knew was they were sufficiently prepared for class. But we were not prepared for what happened next…

Having taught the class many times, our ConcepTests and notes were fairly polished. We entered our class as usual, quite matter of factly. Following the JiTT-Peer Instruction script, we began the class by addressing students' difficulties, briefly lectured students, presented them with a first conceptual question, facilitated peer-discussions and followed up with a simple single-concept problem. To our surprise, students were somewhat irritated by this. Why were we giving a brief lecture on what we had already made them read? And had we not asked them similar kinds of questions before class? Indeed, we had! By tweaking JiTT, we had pushed most of the content outside of class. Whoops, we thought, we just flipped the classroom! Now what should we do?

**Rethinking Our Teaching**

Faced with an unexpectedly flipped classroom, we looked for student-centered active learning activities to do in class: Interactive Lecture Demonstrations[10], more complex problems similar to collaborative group problem solving activities[11, 12]. We were slowly trying to take ownership of our inadvertently flipped classroom.

This educational treatment came with side-effects. Certain things happened that we had not planned for and, in all our years of teaching, we had never seen. First, a student told us that they found E&M less confusing than mechanics. After a severe double-take, we asked the student to clarify. "We're in week 4 now", the student answered. "Last semester in mechanics, I was lost by week 4…" Another surprise came at the end of the semester. Pressed for time, we failed to systematically prepare a thorough warm-up for the last few classes. Before one of these last classes, a few students came asking for the chapters to read before class. In many years of teaching, never had a student approached us asking for chapter sections to read *before a class*. They had often asked us what to read to prepare for a test, but never for a class. However, the most surprising side-effect was one that flipped our own understanding of teaching and learning.

When we realized that we had flipped our classroom, we were not quite sure what to do. So we started by ruling out what we believed we shouldn't do. One thing seemed clear: lecturing was out of the question. We had read the papers (and written a few), attended (and given) talks and workshops on why lecturing just does not work[6, 13, 14]. Then a question arose in class during one of the problem solving sessions. Although we had explicitly acknowledged that we should not lecture, one of us (NL) dove right into a lecture-mode explanation about how the electric field near an infinite sheet of charge was constant when you cannot 'see' the edges, the same way the gravitational field is constant when we are close enough to the Earth that we cannot see its curvature. That is, you can consider "g" as constant if the Earth appears to be flat (there you go; he did it again…). Surprisingly, students were more attentive than ever. Questions and classroom discussions arose. Nowhere in recent memory could we find an instance of this level of engagement in any of our lectures. Could a lecture actually be useful? Upon reflection, we recognized that students might have been more engaged because they had been properly prepared for the lecture.

From our own experiences we know it is possible to have a greater appreciation for academic talks (lectures) when we know more about the subject (and occasionally dozed off when we don't). This idea is not a new one. It was precisely the notion of a well-known learning sciences paper called, A Time-For-Telling[15]. The conclusion remained counter-intuitive to us: *lectures can be useful* **if** students are properly prepared. Our heads are still spinning from that flip…

**Conclusion**

When all the material cannot be covered in class, instructors use JiTT to push part of the course material outside of the classroom. We found that a simple tweak of JiTT, which we've called the flip-JiTT, can easily lead to flipping your classroom because most of the coverage happens before class. We started out looking for active learning methods that would move us away from lectures. Surprisingly, we found that when prepared the right way (e.g. flip-JiTT) students can be engaged by lectures too.